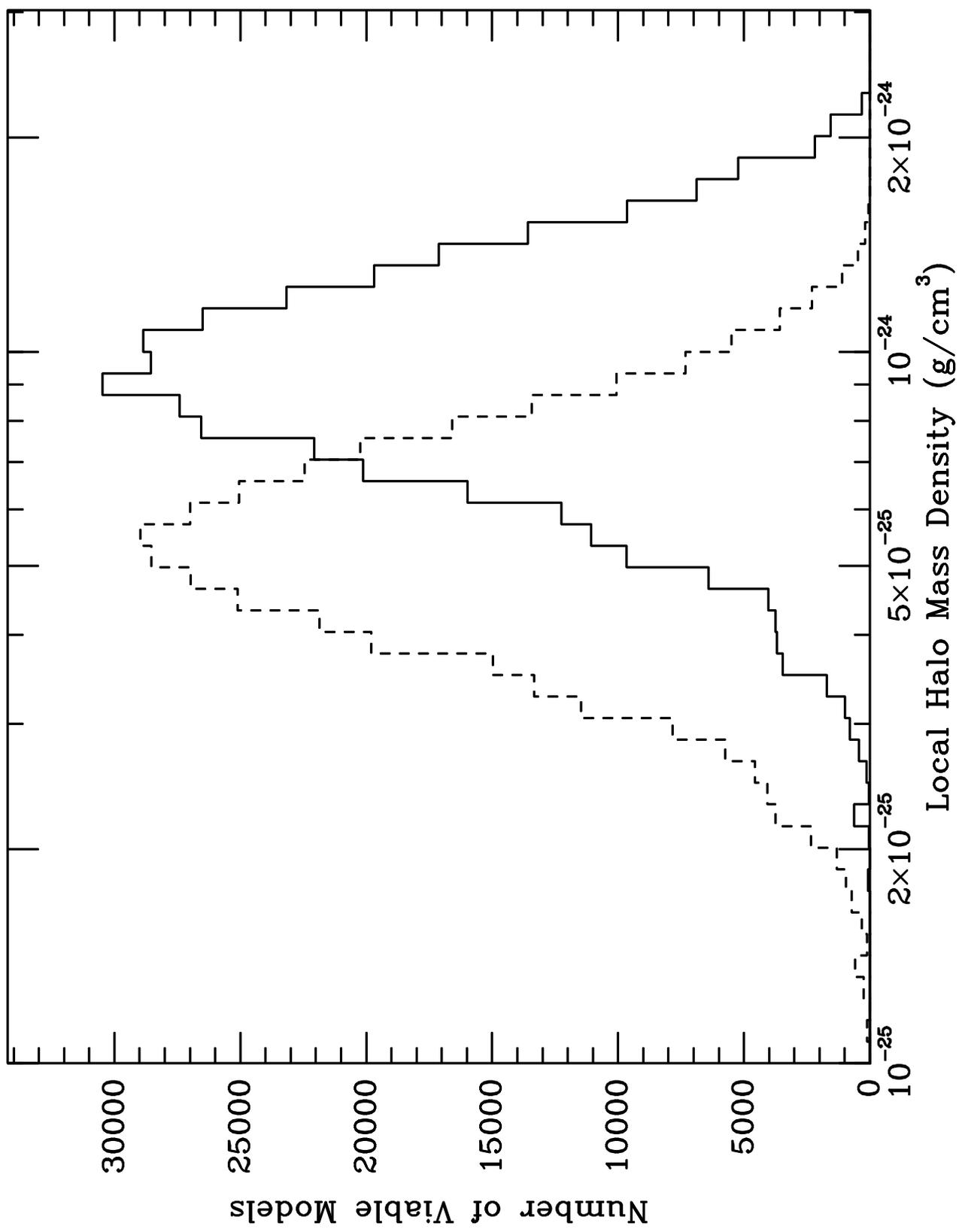

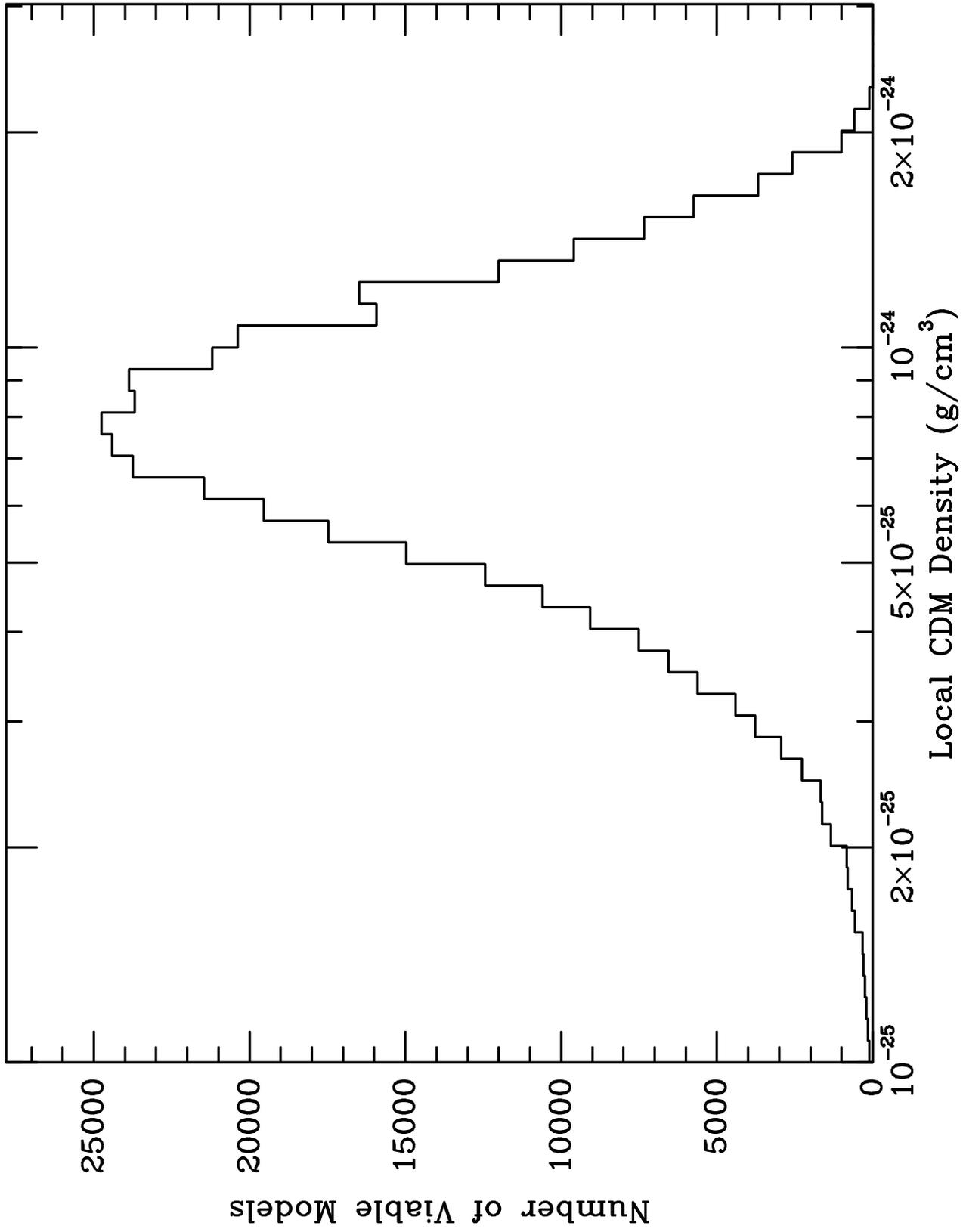

# THE LOCAL HALO DENSITY

Evalyn I. Gates,[1,2] Geza Gyuk,[2,3] and Michael S. Turner[1,2,3]

[1]*Department of Astronomy & Astrophysics*
*Enrico Fermi Institute, The University of Chicago,*
*5640 S. Ellis Ave., Chicago, IL 60637-1433*

[2]*NASA/Fermilab Astrophysics Center*
*Fermi National Accelerator Laboratory, Batavia, IL 60510-0500*

[3]*Department of Physics, Enrico Fermi Institute*
*The University of Chicago,*
*5640 S. Ellis Ave., Chicago, IL 60637-1433*

## ABSTRACT

For almost twenty years models of the Galaxy have included a dark halo responsible for supporting a substantial fraction of the local rotation velocity and a flat rotation curve at large distances. Estimates of the local halo density range from $2 \times 10^{-25}\,\mathrm{g\,cm^{-3}}$ to $10 \times 10^{-25}\,\mathrm{g\,cm^{-3}}$. By careful modeling of the Galaxy, taking account of the evidence that dark halos are flattened and recent microlensing data, we arrive at a more quantitative estimate, $9.2^{+3.8}_{-3.1} \times 10^{-25}\,\mathrm{g\,cm^{-3}}$. Microlensing toward the LMC indicates that only a small fraction, less than about 30%, can be in the form of MACHOs, consistent with the idea that most of the halo consists of cold dark matter particles.

*Subject headings:* dark matter — Galaxy: halo, structure — gravitational lensing

There is an abundance of evidence that spiral galaxies have massive, dark halos; it includes flat rotation curves, warping of the disk and flaring of HI gas, the motions of satellite galaxies (see e.g. Rubin 1993; Knapp & Kormendy 1987; Zaritsky & White 1994; Fich & Tremaine 1991; Persic & Salucci 1995), and most recently evidence for weak-gravitational lensing (Brainerd, Blanford & Smail 1995). However, the uncertainties in the parameters that describe the dark halo—core radius, local density, spatial extent, and shape—are large because observations do not tightly constrain it. To illustrate, estimates of the local halo density range from about $2 \times 10^{-25}\,\mathrm{g\,cm^{-3}}$ to $10^{-24}\,\mathrm{g\,cm^{-3}}$ (Caldwell & Ostriker 1981; Bahcall, Schmidt & Soneira 1983; Schmidt 1985; Turner 1986; Flores 1988; Freeman 1987). For reference, $0.01 M_\odot\,\mathrm{pc}^{-3} = 6.8 \times 10^{-25}\,\mathrm{g\,cm^{-3}} = 0.38\,\mathrm{GeV\,cm^{-3}}$.

The possibility that the dark halos of spiral galaxies are comprised mostly of slowly moving elementary particles (referred to as cold dark matter (see e.g. Turner 1993)) has spurred further interest in the properties of the dark halo of the Galaxy, especially the local density. This is because several large-scale efforts are underway to directly detect cold dark matter particles in our neighborhood: neutralinos by their elastic scattering in kg-mass detectors (Primack, Seckel & Sadoulet 1988; Smith & Lewin 1990) and axions by their conversion to microwave photons in strong magnetic fields (van Bibber et al. 1992). In addition, other indirect methods are being pursued, e.g., the search for high-energy neutrinos from neutralino annihilations within the Sun and Earth, and the search for cosmic-ray positrons, gamma rays and antiprotons from neutralino annihilations in the halo (Jungman, Kamionkowski & Griest 1995). Here too the properties of the dark halo are important.

There is circumstantial evidence that the halos of spiral galaxies are comprised primarily of cold dark matter particles. First, there is the failure to find enough dwarf stars in searches for faint objects (see e.g., Bahcall et al. 1994) or MACHOs in microlensing searches (see e.g., Gates, Gyuk & Turner 1995a) to account for but a small fraction of the halo. (It is still possible that the dark halo is comprised of clouds of molecular hydrogen whose distribution on the sky is a fractal (De Paolis et al. 1995) or $10^3 M_\odot$ black holes.) Likewise, the mass fraction of rich clusters that can be readily identified as baryons, overwhelmingly in the form of hot, x-ray emitting gas, is small (see e.g., White et al. 1993). Finally, and perhaps most importantly, the most successful models of structure formation involve cold dark matter; e.g., the various versions of "the cold dark matter theory" (cosmological constant + cold dark matter, hot + cold dark matter, unstable tau neutrino + cold dark matter, tilted cold dark matter, and cold dark matter + a low value of the Hubble constant) (see e.g. Liddle & Lyth 1993; Bartlett et al. 1995; Dodelson, Gyuk & Turner 1994; Primack et al. 1995), as well as cold dark matter with topological defects such as textures (Turok 1991). The last two pieces of circumstantial evidence only directly support that the idea that the bulk of the matter in the Universe is cold dark matter; because cold dark matter particles move so slowly, they inevitably must find their way into galactic halos including our own (see e.g., Gates & Turner 1994).

The purpose of this *Letter* is to quantify the local mass density of halo material. Our approach is very similar to that used by Caldwell and Ostriker (1981) almost fifteen years ago. We construct models of the Galaxy, compare them to the observational data, and then examine the local halo density in "viable" models, that is, those models that are consistent with the observational data. In this regard we have added two additional important pieces of data: microlensing and evidence that halos are flattened significantly. We have tried to be more quantitative than previous studies, and we have displayed our results in plots that *resemble* likelihood functions that are marginalized with respect to the local halo mass density, cf. Figs. 1 and 2. In fact, they are not likelihood distributions; rather, they are histograms of the number of viable models as a function of local halo density. Because the most important uncertainties in modelling the Galaxy are systematic in character—the model of the Galaxy itself, the rotation curve, the shape of the halo, and even the galactocentric distance and local speed of rotation—we resisted the urge to carry out a more rigorous statistical analysis which might have conveyed a false level of statistical significance.

Our methodology is described in detail elsewhere (Gates, Gyuk & Turner 1995b); here we briefly review it. We begin with a model of the Galaxy that consists of: (1) a bar or an asymmetric bulge; (2) a disk with a fixed luminous component and a variable dark lensing component; (3) isothermal cold dark matter and MACHO halos with finite core radii and ellipsoidal density contours,

$$\rho_i(R,z) \propto \frac{1}{a_i^2 + R^2 + z^2/q^2}, \qquad (1)$$

where $i$ = MACHOs or cold dark matter and $q^{-1}$ describes the flattening of the halo.[1] The parameters that describe each component are allowed to vary over a "plausible" range: (1) Dwek (1995) and Kent (1992) bulge models with masses of $1 \times 10^{10} M_\odot$ to $4 \times 10^{10} M_\odot$ (see e.g., Blum 1995); (2) in addition to a fixed luminous disk component with scale height 300 pc, scale length 3.5 kpc (Bahcall & Soneira 1980), and local projected surface mass density[2] of $25 M_\odot\,\mathrm{pc}^{-2}$, a double-exponential dark disk with a minimum local projected surface mass density of $10 M_\odot\,\mathrm{pc}^{-2}$, scale heights of 300 pc or 1.5 kpc, scale lengths of 2.5 kpc to 4.5 kpc or a dark Mestel model (surface density varying as $r^{-1}$); (3) independent MACHO and cold dark matter halos with core radii of 2 kpc to 20 kpc, either spherical ($q=1$) or flattened by 2.5 to 1 ($q=0.4$).

From the tens of millions of models we construct, we identify viable models by requiring that they satisfy the following constraints: (a) galactocentric distance of 7.0 kpc to 9.0 kpc (Schechter 1994); (b) local rotation speed of $200\,\mathrm{km\,s^{-1}}$ to $240\,\mathrm{km\,s^{-1}}$; (c) rotation curve whose peak to trough variation between 4 kpc and 18 kpc is no greater than 14% (flat-rotation curve constraint); (d) rotation speed at 60 kpc that is at least 150 km s$^{-1}$ but less than 307 km s$^{-1}$; (e) local projected mass density within 1.0 kpc of the plane (summed over all components) that is less than 100 $M_\odot$ pc$^{-2}$ (see e.g. Bahcall, Flynn & Gould 1992; Kuijken & Gilmore 1991); (f) optical depth for microlensing toward Baade's window in the galactic bulge that is greater than $2 \times 10^{-6}$ (Udalski et al. 1994a,b; Alcock et al. 1995b).

---

[1] We do not consider models with a "tilt" between the disk and halo symmetry axes. Not only does there seem to be little motivation for such, but tilt does not have long-term stability. In any case, for large tilt angle, the optical depth toward the LMC could be strongly enhanced or suppressed (Frieman & Scoccimarro 1994). Even so, we expect our results for the total halo density to be essentially the same.

[2] Estimates of the local projected mass density within 1 kpc of the plane range from $49 M_\odot\,\mathrm{pc}^{-2}$ (Bahcall, Flynn & Gould 1992; Kuijken & Gilmore 1991) to a more recent determination of $39 M_\odot\,\mathrm{pc}^{-2}$ (Gould, Bahcall & Flynn 1995). We choose $25 M_\odot\,\mathrm{pc}^{-2}$ as an estimate of that portion of the luminous disk that does not contribute to microlensing (i.e. gas and bright stars)

In Fig. 1 (dashed line) we show the distribution of the local halo mass density in viable models that have a spherical halo. The distribution peaks at a mass density of $5.3 \times 10^{-25}$ g cm$^{-3}$ and at full-width, half-maximum varies from $3.3 \times 10^{-25}$ g cm$^{-3}$ to $7.5 \times 10^{-25}$ g cm$^{-3}$. Such a range for the local halo density is fully consistent with previous estimates, whose range was estimated less quantitatively. The distribution is most sensitive to the local rotation velocity and the local projected mass density.

Specifically, the mass interior to the solar circle that is required to support the local rotation velocity scales roughly as velocity squared; larger values for the local rotation speed thus systematically lead to higher values for the local halo density. Likewise, for a given local rotation speed, decreasing the local projected mass density leads to less mass in the disk and therefore the necessity for more mass in the halo component and a larger value of the local halo density. (For example, requiring $\Sigma_0 \leq 50 M_\odot$ pc$^{-2}$ rather than $100 M_\odot$ pc$^{-2}$ increases the central value of the halo mass density from $5.3 \times 10^{-25}$ g cm$^{-3}$ to more than $6 \times 10^{-25}$ g cm$^{-3}$.)

The optical depth for microlensing toward the bulge provides an important constraint to the halo. If we raise the minimum acceptable optical depth, then the bulge mass is pushed upward. In turn, this pushes the disk mass downward, as the bulge accounts for more of the inner rotation curve, and the halo mass up to explain the local rotation velocity. The central value for the optical depth to the bulge is $3 \times 10^{-6}$ (perhaps even higher if efficiencies are lower than current estimates (Bennett et al. 1994)), and thus it is entirely possible that with more data the local halo mass density will, for this reason, be pushed upward.

Next, we consider the effect of flattening the halo. The shapes of the dark halos of spiral galaxies are presently not well determined by observational data. Rix (1995) has recently reviewed this issue and has concluded that the two most revealing pieces of evidence are the rotation curves of polar-ring galaxies and numerical simulations of the formation of galactic halos and disks. In polar-ring galaxies one is able to obtain rotation curves both in and perpendicular to the galactic plane. Halo models from E5 to E8 can account for the polar-ring galaxy rotation curves; they are best fit with a flattening of 2.5 to 1 (E6) (Sackett et al. 1994). Numerical simulations indicate that halos formed from dissipationless matter are triaxial with a distribution of flattening that peaks around 1.4 to 1, and further, that the subsequent formation of a disk tends to reduce the triaxiality and to increase the amount of halo flattening. To summarize, there is no reason to believe that halos are spherical and the data that exist indicate that an E6 model provides a good description.

As we shall describe, the effects of flattening on the local halo density scale in a simple way, and so we shall focus on an E6 halo model, where the halo is flattened by 2.5 to 1 ($q = 0.4$). The distribution of total local halo mass density (baryons + MACHOs) for this model and for comparison, a spherical model, are shown in Fig. 1. Note that the distributions are almost identical, except that the local halo density in the flattened model is about a factor of two larger.

This enhancement can be easily understood and its magnitude accurately estimated. Consider a spherical halo; now imagine flattening it by a factor $q^{-1}$ while holding the mass fixed. The halo mass density in the equatorial plane increases by a factor of $q^{-1}$. The force exerted on a test particle moving in the equatorial plane is still only due to the mass elements inside the now flattened density contour that runs through its position (Binney & Tremaine 1987), and it is increased only slightly due to geometrical factors. In order to keep the local rotation velocity fixed the halo mass density must be reduced by a small factor (whose value reaches a maximum of $\pi/2$ in the limit $q \to 0$); the net effect is an increase in the local density of slightly less than $q^{-1}$.

One can be more quantitative; comparing a spherical halo model with a flattened model with the same asymptotic rotation velocity and core radius, one finds the halo density in the equatorial plane is larger for the flattened model by a factor

$$f = \frac{\sqrt{1-q^2}}{q \sin^{-1}(\sqrt{1-q^2})}, \qquad (2)$$

and in the limit $q \ll 1$, $f \to 2/\pi q$. (This equation follows directly from the equation for the equatorial rotation curve produced by a flattened halo, which is given in Binney & Tremaine (1987).) For a flattening of 2.5 ($q = 0.4$), an enhancement in the local halo mass density of about 1.98 is expected, which is consistent with the shift seen in Fig. 1.

Finally, a few more comments concerning microlensing. Microlensing has provided and will continue to provide an important new probe of the Galaxy (Paczynski 1986). Thus far two directions have been explored: a line of sight toward the LMC which mainly probes the halo and several lines of the sight toward the bulge which mainly probe the inner Galaxy. The higher than expected optical depth for microlensing toward the bulge has provided additional, very strong evidence that the bulge more closely resembles a bar (Blitz 1993; Binney 1993; Dwek 1995; Stanek et al. 1994; Zhao, Spergel & Rich 1994). Further, as mentioned above, by constraining the inner Galaxy it indirectly constrains the halo.

The smaller than expected optical depth for microlensing toward the LMC (Aubourg et al. 1993; Alcock et al. 1995) has provided strong evidence against an all-MACHO halo, and suggests that the mass fraction of the halo that is in MACHOs is less than 30% (Gates et al. 1995a). By imposing a further constraint based upon microlensing toward the LMC we can obtain the distribution for the local cold dark matter density. For viable models we require an optical depth for microlensing toward the LMC that is between $0.2 \times 10^{-7}$ and $2 \times 10^{-7}$. In Fig. 2 we show the distribution of the local density of cold dark matter particles (in a model with a halo flattening of 2.5 to 1); it is very similar to Fig. 1, but shifted to lower densities, by a factor of about 0.8.

As an aside, we emphasize that it is not yet possible to rule out a halo that is comprised almost entirely of MACHOs. We have found a few viable models (which incorporate the above LMC microlensing constraint) where the halo MACHO fraction is greater than 75% (Gates et al. 1995a). These models are characterized by a very "light" halo; more specifically, a heavy dark disk ($\Sigma_{\text{dark disk}} \gtrsim 40 M_\odot \, \text{pc}^{-2}$), small local rotation speed (less than $220 \, \text{km s}^{-1}$), small asymptotic rotation speed (around $150 \, \text{km s}^{-1}$), low escape velocity ($\lesssim 450 \, \text{km s}^{-1}$), high microlensing rate toward the LMC (around $2 \times 10^{-7}$), and small microlensing rate toward the bulge (around $2 \times 10^{-6}$). Because of these distinctive features, the possibility of an all-MACHO halo can be falsified by any of a number of measurements: bulge microlensing optical depth larger than about $3 \times 10^{-6}$, LMC microlensing optical depth less than about $1.5 \times 10^{-7}$, determination that the total projected mass density is less than about $60 M_\odot \, \text{pc}^{-2}$, conclusive evidence that the local escape velocity is $450 \, \text{km s}^{-1}$ or larger, and measurement of the asymptotic value of the rotation velocity of the Galaxy that is greater than around $180 \, \text{km s}^{-1}$. This is discussed in much greater detail in Gates, Gyuk & Turner (1995b).

In summary, we have quantified expectations for the local mass density of the dark halo of the Galaxy. We find that in most models of the Galaxy that have an E6 halo and are consistent with observational data, the local halo mass density is between $6.1 \times 10^{-25} \, \text{g cm}^{-3}$ and $13 \times 10^{-25} \, \text{g cm}^{-3}$. If the halo were more spherical, say E5, the local halo mass density would be about 15% smaller, and if it were flatter, say E8, it would be about 1.8 times greater. [In general the local halo mass density scales according to Eq. (2)]. Searches for faint dwarf stars and dark stars (MACHOs) indicate they can account for only a fraction of the halo material (less than 30%). If the bulk of the unexplained halo material is cold dark matter, its local density is between about $4.5 \times 10^{-25} \, \text{g cm}^{-3}$ and $12 \times 10^{-25} \, \text{g cm}^{-3}$, a factor of almost two higher than previous estimates, which is welcome news for those involved in the search for particle dark matter.

# Acknowledgments

This work was supported in part by the DOE (at Chicago and Fermilab) and the NASA (at Fermilab through grant NAG 5-2788).

# Figure Captions

**Figure 1:** Distribution of the total local halo density in viable models of the Galaxy with spherical (broken) and E6 (solid) halos.

**Figure 2:** Distribution of the local cold dark matter mass density in viable models of the Galaxy with flattened halos (2.5 to 1).